\documentclass[aps,showkeys,twocolumn,preprintnumbers,showpacs,amsmath,amssymb,superscriptaddress,floatfix,nofootinbib]{revtex4}
\usepackage{graphicx,color,dcolumn,booktabs,bm}
\usepackage{longtable,lscape}
\usepackage{txfonts}
\usepackage{overpic}
\usepackage{epsfig}
\usepackage{amssymb}
\usepackage{rotating}
\usepackage{epstopdf}
\usepackage{appendix}
\usepackage{indentfirst}
\usepackage{feynmf}   
\usepackage{slashed}  
\usepackage{cases}
\usepackage{color}
\usepackage{multirow}
\usepackage{graphicx,color,dcolumn,booktabs,bm}
\usepackage{cases}
\usepackage{array}

\graphicspath{{Figures/}} %

\usepackage[colorlinks, citecolor=blue,anchorcolor=red,menucolor=red, linkcolor=red,filecolor=red,runcolor=red,urlcolor=blue,frenchlinks=red]{hyperref}

\newcommand{\DKM}{{^3\!P_0}}
\newcommand{\bsp}{{\boldsymbol{p}}}
\newcommand{\Ycal}{{\cal{Y}}}
\newcommand{\Mcal}{{\cal{M}}}
\newcommand{\ddd}{{\rm{d}^3}}

\begin{document}
\title{Canonical interpretation of the $D_{s0}(2590)^{+}$ resonance  }

\author{Zhuo Gao}
\affiliation{School of Physics and Electronics, Henan University, Kaifeng 475004, China}

\author{Guan-Ying Wang\footnote{Corresponding author }}
\email{wangguanying@henu.edu.cn}
\affiliation{School of Physics and Electronics, Henan University, Kaifeng 475004, China}

\author{Qi-Fang L\"u\footnote{Corresponding author }}
\email{lvqifang@hunnu.edu.cn} %
\affiliation{  Department
	of Physics, Hunan Normal University,  Changsha 410081, China }

\affiliation{ Synergetic Innovation
	Center for Quantum Effects and Applications (SICQEA), Changsha 410081,China}

\affiliation{  Key Laboratory of
	Low-Dimensional Quantum Structures and Quantum Control of Ministry
	of Education, Changsha 410081, China}

\author{Jingya Zhu}
\affiliation{School of Physics and Electronics, Henan University, Kaifeng 475004, China}

\author{Gao-Feng Zhao\footnote{Corresponding author }}
\email{10110094@vip.henu.edu.cn}
\affiliation{School of Physics and Electronics, Henan University, Kaifeng 475004, China}

\date{\today}

\begin{abstract}
	
The $D_{s0}(2590)^{+}$ resonance observed by LHCb Collaboration is a strong candidate of the $D_{s}(2^1S_0)$ state according to its spin parity and strong decay mode. However, the measured mass seems relatively lower than the previous theoretical predictions, which interests the coupled channel interpretations in the literature. In this work, we adopt an alternate approach, taking into account the  screening effects in the potential model, to describe the  $D_{s0}(2590)^{+}$ resonance.  The mass spectrum and strong decays of the excited charmed-strange mesons are investigated within the modified relativized quark model and $^3P_0$ model. The calculated mass and width of the $D_{s0}(2590)^{+}$ are consistent with the experimental observations, which indicate that it can be reasonably interpreted as the $D_{s}(2^{1}S_{0})$ state.
\end{abstract}
\pacs{ }
\maketitle

\section{Introduction}{\label{Introduction}}

A heavy-light meson is composed of one heavy quark and one light antiquark, and acts as a hydrogen atom.  Understanding the heavy-light meson spectrum and searching
for the missing resonances are important tasks in hadronic physics, which provide us a good opportunity to deepen our understanding of the complicated non-perturbative behavior of QCD in the low energy regime. Among of them, the charmed-strange sector is particularly interesting and has gained wide attentions, since the mysterious $D_{s0}^\ast(2317)$ and $D_{s1}(2460)$ states observed by BaBar and CLEO Collaborations ~\cite{BaBar:2003oey,CLEO:2003ggt} have rather lower masses compared with the theoretical predictions in conventional quark models~\cite{Godfrey:2015dva,Ebert:2009ua,Zeng:1994vj,Lahde:1999ih,DiPierro:2001dwf,Li:2010vx,Ni:2021pce}.

From the Review of Particle Physics~\cite{ParticleDataGroup:2021}, there exist twelve states in the charmed-strange sector. The $D_s$ and $D_s^*$ are the ground states, and $D_{s1}(2536)$ and $D_{s2}^*(2573)$ can be well understood as the $P-$wave states. The $D_{s1}^*(2700)$, $D_{s1}^*(2860)$, and $D_{s3}^*(2860)$ can be assigned as the $D_s(2^3S_1)$, $D_s(1^3D_1)$, and $D_s(1^3D_3)$ states, respectively, where the $2S-1D$ mixing effect may be also significant for the vector mesons.  The canonical interpretations of $D_{s0}^\ast(2317)$ and $D_{s1}(2460)$ are problematic in the traditional quark model, while the $X_0(2900)$ and $X_1(2900)$ observed by LHCb Collaboration are undoubtedly exotic~\cite{LHCb:2020pxc,LHCb:2020bls}. Moreover, information on the $D_{sJ}(3040)$ state is quite limited, which prevents us to reach a definite conclusion. It can be seen that the low-lying charmed-strange spectrum is far from being established.

Recently, the LHCb Collaboration observed a new excited resonance $D_{s0}(2590)^{+}$  in the $D^{+} K^{+}\pi^{-}$ mass distribution of the $B^{0}\rightarrow D^{-}D^{+}K^{+}\pi^{-}$ decay~\cite{LHCb:2020gnv}. Its mass, width, and spin parity are determined to be $m=2591\pm6\pm7$ MeV, $\Gamma=89\pm16\pm12$ MeV, and $J^{P}=0^{-}$, respectively. Based on these properties, the LHCb Collaboration suggested that this state is a strong candidate for the radial excited $D_{s}(2^{1}S_{0})$ state. However, the measured mass seems relatively lower than the previous theoretical predictions in the literature~\cite{Godfrey:2015dva,Ebert:2009ua,Zeng:1994vj,Lahde:1999ih,DiPierro:2001dwf,Li:2010vx,Ni:2021pce,Godfrey:1985xj}, which leads to different interpretations on the theoretical side. In Ref.~\cite{Wang:2021orp}, the authors investigated the mass and width of  $D_{s0}(2590)^+$ by solving the Bethe-Salpeter equation, and concluded that it can be hardly explained as $D_s(2^1S_0)$ state. Within the semi-relativistic potential model and chiral quark model, the mass and width of $D_{s0}(2590)^+$ are not consistent with that of $D_s(2^1S_0)$ state~\cite{Ni:2021pce}. In Ref.~\cite{Xie:2021dwe}, the authors employ the unquenched quark model to describe the mass of $D_{s0}(2590)^+$ by considering the mixture of the $D_s(2^1S_0)$ state and $D^\ast K$ component. Also, the authors performed a coupled-channel calculation including the $D^{(\ast)}K^{(\ast)}$, $D_s^{(\ast)}\omega$ and $D_s^{(\ast)}\eta$ channels, and found that the $D_{s0}(2590)$ can be regarded as a bare $D_s(2^1S_0)$ state plus dominant $D^\ast K$ part~\cite{Ortega:2021fem}.  These theoretical works suggest that the $D_{s0}(2590)^+$ may be not a pure $D_s(2^1S_0)$ state and the $D^\ast K$ component should be significant.

 Based on the $SU(3)$ light quark flavor symmetry, the $P-$wave charmed-strange mesons is supposed to be higher than their charmed partners. The violation of $SU(3)$ flavor symmetry for $D_{s0}^\ast(2317)$ and $D_{s1}(2460)$ resonances suggest that they are not pure $P-$wave $c \bar s$ states and the coupled-channel effects are essential. Actually, in the literature, the coupled-channel approach, meson-loop effects, or the unquenched quark model has been widely discussed in the charmed-strange sector to reduce the theoretically predicted masses~\cite{Xie:2021dwe,Ortega:2021fem,Tan:2021bvl,Browder:2003fk, Hwang:2005tm, Lu:2006ry, Bicudo:2005de, Mohler:2011ke, MartinezTorres:2011pr, Mohler:2013rwa, Lang:2014yfa, Ortega:2016mms, MartinezTorres:2017bdo, Bali:2017pdv, Albaladejo:2018mhb}, which mainly focused on the $D_{s0}^\ast(2317)$ and $D_{s1}(2460)$ resonances. However, the situation of the $D_{s0}(2590)^+$ is better. The mass gap between  $D_{s1}^*(2700)$ and $D_{s0}(2590)$ is
\begin{eqnarray}
	m[D_{s1}^*(2700)] - m[D_{s0}(2590)] = 123~\rm{MeV},
\end{eqnarray}
and the mass gap between two $2S$ charmed states $D_1^*(2600)$ and $D_0(2550)$ with the latest measurements of LHCb Collaboration is~\cite{LHCb:2019juy}
\begin{eqnarray}
	m[D_1^*(2600)] - m[D_0(2550)] = 124~\rm{MeV}.
\end{eqnarray}
The approximately equal mass splittings of charmed and charmed-strange  sectors strongly suggest that the $D_{s0}(2590)$ should be the partner of $D_0(2550)$ and can be assigned as the $D_s(2^1S_0)$ state as the LHCb Collaboration suggested.

Instead of the unquenched approaches with higher Fock states, the potential model including screening effects is an  alternate approach to lower the mass spectrum, which has been extensively employed to study the properties of conventional mesons and achieved significant success. The advantage of the screening potential is that one can bring down the masses of excited states while avoiding involving higher Fock components. Hence, we expect that the potential model including screening effects may relieve the tension between measured mass and theoretical predictions under the assignment of $D_{s0}(2590)$ as $D_s(2^1S_0)$ state.  Moreover, it is natural and necessary to explore the possible conventional descriptions for a newly observed particle before introducing more complicated and exotic configurations. In this work, we apply the Godfrey-Isgur's relativized quark model including screening effects to revisit the mass spectrum of the charmed-strange mesons, and then adopt the obtained wave functions to study their strong decay behaviors in the $\DKM $ model. Our results show that the calculated mass and width of the $D_{s0}(2590)^{+}$ are consistent with the experimental observations, which suggest that it can be reasonably interpreted as the $D_{s}(2^{1}S_{0})$ state.

This article is organized as follows. In Sec.~\ref{sec:formalisms}, we briefly introduce the relativized quark models and $\DKM$ models. The results and discussions of charmed-strange mesons are presented in Sec.~\ref{sec:result}. Finally, a summary is given in the last section.

\section{Models}
\label{sec:formalisms}

\subsection{The relativized quark model}
\label{sec:GI}

In this subsection, we will give a brief introduction of the relativized quark model proposed by Godfrey and Isgur (GI model)~\cite{Godfrey:1985xj}. This model has been extensively adopted to investigate the properties of conventional hadrons~\cite{Godfrey:1985xj,Godfrey:2014fga, Godfrey:2015dva, Li:2021qod, Godfrey:2004ya, Capstick:1985xss,Barnes:2005pb,Sun:2014wea,Godfrey:2016nwn} and tetraquarks~\cite{Lu:2021kut, Lu:2020cns, Lu:2020rog, Lu:2020qmp, Lu:2016cwr, Lu:2016zhe, Lu:2019ira, Anwar:2017toa, Anwar:2018sol, Bedolla:2019zwg, Ferretti:2020ewe}, and give a unified description of different flavor sectors. In particular, for the low-lying states, the relativized quark model plays an important role in studying their mass spectra and provides an effective criterion to distinguish conventional mesons from exotics.

For a two-body system, the relevant Hamiltonian can be written
as
\begin{equation}
	H = H_0+V^{\rm oge}+V^{\rm conf}, \label{ham}
\end{equation}
where
\begin{equation}
	H_0 = \sqrt{p^2+m_1^2}+\sqrt{p^2+m_2^2}
\end{equation}
is the relativistic kinetic energy, $V^{\rm oge}$ is the one gluon exchange potential, and $V^{\rm conf}$ corresponds to the confining potential. The induced spin-dependent interactions are also included in the $V^{\rm oge}$ and $V^{\rm conf}$.

More explicitly,  the potentials $V^{\rm oge}$ and $V^{\rm conf}$ can be expressed as
\begin{eqnarray}
	V^{\rm{oge}} &=& \beta_{12}^{1/2}\tilde G(r)\beta_{12}^{1/2}  +\delta_{12}^{1/2+\epsilon_{\rm c}} \frac{2\boldsymbol{S_1}
		\cdot \boldsymbol{S_2}}{3m_1m_2} \nabla^2\tilde G(r) \delta_{12}^{1/2+\epsilon_{\rm c}}\nonumber \\  && +\delta_{12}^{1/2+so(v)} \frac{(\bm {S_1}+\bm {S_2}) \cdot \bm L}{m_1m_2} \frac{1}{r}\frac{\partial \tilde G(r)}{\partial r}\delta_{12}^{1/2+so(v)}  \nonumber \\  &&  + \delta_{11}^{1/2+so(v)}\frac{\bm {S_1} \cdot \bm L}{2m_1^2} \frac{1}{r}\frac{\partial \tilde G(r)}{\partial r}\delta_{11}^{1/2+so(v)}   \nonumber \\  &&+ \delta_{22}^{1/2+so(v)}\frac{\bm {S_2} \cdot \bm L}{2m_2^2} \frac{1}{r}\frac{\partial \tilde G(2)}{\partial r}\delta_{22}^{1/2+so(v)}  \nonumber \\  && +\delta_{12}^{1/2+\epsilon_{\rm t}}\Bigg(\frac{\bm{S_1}\cdot \hat {\bm r} \bm{S_2}\cdot \hat {\bm r}-\bm{S_1} \cdot \bm{S_2}/3}{m_1m_2} \Bigg) \nonumber \\  && \times \Bigg( \frac{1}{r}\frac{\partial}{\partial r}-\frac{\partial^2}{\partial r^2}\Bigg)\tilde G(r)\delta_{12}^{1/2+\epsilon_{\rm t}},
\end{eqnarray}
and
\begin{eqnarray}
	V^{\rm{conf}} &=& \tilde S(r) - \delta_{11}^{1/2+so(s)}\frac{\bm {S_1} \cdot \bm L}{2m_1^2} \frac{1}{r}\frac{\partial \tilde S(r)}{\partial r}\delta_{11}^{1/2+so(s)}  \nonumber \\  && - \delta_{22}^{1/2+so(s)}\frac{\bm {S_2} \cdot \bm L}{2m_2^2} \frac{1}{r}\frac{\partial \tilde S}{\partial r}\delta_{22}^{1/2+so(s)}.
\end{eqnarray}
Here, the $\tilde G(r)$ and $\tilde S(r)$ are the smeared potentials, and can be written as
\begin{equation}
	\tilde G(r) = - \sum_{k=1}^3\frac{4\alpha_k}{3r}{\rm erf}(\tau_{k12}r)
\end{equation}
and
\begin{eqnarray}
	 \tilde S(r) &=& br \left[\frac{e^{-\sigma_{12}^2r^2}}{\sqrt{\pi}\sigma_{12}r}+\left(1+\frac{1}{2\sigma_{12}^2r^2}\right)
	{\rm erf}(\sigma_{12}r)\right]+c
\end{eqnarray}
with
\begin{equation}
	\frac{1}{\tau_{k12}^2} = \frac{1}{\gamma_k^2}+\frac{1}{\sigma_{12}^2}
\end{equation}
and
\begin{equation}
	\sigma_{12}^2 = \sigma_0^2
	\left[\frac{1}{2}+\frac{1}{2}\left(\frac{4m_1m_2}{(m_1+m_2)^2}\right)^4\right]+s^2\left(\frac{2m_1m_2}{m_1+m_2}\right)^2.
\end{equation}
The definition of $\delta_{11}$, $\delta_{12}$, $\delta_{22}$, and $\beta_{12}$ are
\begin{equation}
	\delta_{ij} = \frac{m_im_j}{(p^2+m_i^2)^{1/2}(p^2+m_j^2)^{1/2}}
\end{equation}
and
\begin{equation}
	\beta_{12} = 1+\frac{p^2}{(p^2+m_1^2)^{1/2}(p^2+m_2^2)^{1/2}}.
\end{equation}
The $p$ is the magnitude of the relative momentum between the quark and antiquark. The $m_1$ and $m_2$ are masses of the quark and antiquark, respectively. The $\alpha_k$, $\gamma_k$, $b$, $c$, $\sigma_0$, $s$ and  $\epsilon_i$ are the parameters introduced in the relativized quark model.

\subsection{Screened potential}
\label{sec:Screened}

For high excited states, it is necessary to introduce the screening effects to the relativized model, because the linear confining potential will be screened and softened by the vacuum polarization effects at a large distance~\cite{Li:2009zu,Chao:1992et, Ding:1993uy}. Also, the modified relativized model (MGI model) including screen effects turns out to be able to give a better description of the mass spectra for the radial and orbital excitations~\cite{Song:2015fha,Song:2015nia,Lu:2016cwr,Wang:2018rjg,Pang:2017dlw,Pang:2018gcn, Pang:2019ttv,Hao:2019fjg}.

To incorporate the screen effects in the relatived quark model, we should replace the confining potential $\tilde S(r)$ with a screened potential. The $\tilde S(r)$ actually arises from the liner confinement according to the smearing transformation. For an arbitrary potential $f(r)$, the smeared ones $\tilde f(r)$ can be expressed as
\begin{eqnarray}\label{sme}
	\tilde{f}(r)=\int d^3r'\rho_{12}(\mathbf{r}-\mathbf{r'})f(r')
\end{eqnarray}
with
\begin{eqnarray}
	& \rho_{12}\left(\mathbf{r}-\mathbf{r'}\right)=\frac{\sigma_{12}^3}{\pi ^{3/2}}e^{-\sigma_{12}^2\left(\mathbf{r}-\mathbf{r'}\right)^2}.
	\label{si12}
\end{eqnarray}
It can be noticed that the linear confining potential $S(r)=br+c$ indeed leads to the $\tilde S(r)$ through the above smearing transformation. Here, the constant $c$ always attaches to the confining potential for the same convention as Ref.~\cite{Godfrey:1985xj}, which can be fixed by the mass of the ground state.

In the literature, the following replacement is often employed to modify the linear confining potential in the quark model \cite{Chao:1992et, Ding:1993uy},
\begin{eqnarray}
	S(r)=br+c \rightarrow V^{scr}(r)=\frac{b(1-e^{-\mu r})}{\mu}+c.
\end{eqnarray}
If $r$ is small enough, one has $V^{scr}(r)=V(r)$. Therefore, this replacement will minimally affect the ground states, and reduce the excited states significantly. The parameter $\mu$ is related to the strength of the screening effects, and one can roughly understand that the screening effects begin to work from $r \sim 1/\mu$. With the smearing transformation, one have
\begin{eqnarray}
	\tilde V^{scr}(r)&=&\frac{b}{\mu r}\Bigg[e^{\frac{\mu^2}{4 \sigma_{12}^2}+\mu r}\Bigg(\frac{1}{\sqrt{\pi}}\int_0^{\frac{\mu+2r\sigma_{12}^2}{2\sigma_{12}}}e^{-x^2}dx-\frac{1}{2}\Bigg)\nonumber\\
	&&\times\frac{\mu+2r\sigma_{12}^2}{2\sigma_{12}^2}+r-e^{\frac{\mu^2}{4\sigma_{12}^2}-\mu r}\frac{\mu-2r\sigma_{12}^2}{2\sigma_{12}^2}\nonumber\\
	&&\times\Bigg(\frac{1}{\sqrt{\pi}}\int_0^{\frac{\mu-2r\sigma_{12}^2}{2\sigma_{12}}}e^{-x^2}dx-\frac{1}{2}\Bigg)\Bigg]+c. \label{Eq:pot}
\end{eqnarray}

Finally, by replacing the $\tilde S(r)$ with $\tilde V^{scr}$ in the original relativized quark model, we obtain the modified relativized quark model including the screening effects. The mass spectrum and wave functions of the mesons can be obtained by solving the relativized Hamiltonian, and the wave functions are used as inputs to investigate the subsequent strong decays for mesons.

\subsection{ The $\DKM$ model }
\label{sec:3p0}
 In addition to the mass spectrum, the decay widths are crucial  to identify the assignments for mesons.  Here, we give a brief introduction of the $\DKM$ model which is widely used in studying two-body OZI-allowed strong decays of mesons~\cite{Barnes:2002mu,Wang:2017pxm,Roberts:1992js,Blundell:1996as, Barnes:1996ff, Close:2005se, Barnes:2005pb, Zhang:2006yj,
Ding:2007pc, Li:2008mza, Li:2008we, Li:2008et, Li:2008xy, Li:2009rka, Li:2009qu, Li:2010vx,Lu:2014zua, Pan:2016bac,Lu:2016bbk}. In the $\DKM$ model, the strong decay of a meson takes place by producing a quark-antiquark pair with vacuum quantum number $J^{PC}=0^{++}$. The newly created quark-antiquark pair, together with the $q\bar{q}$ in the initial meson, regroups into two outgoing mesons in all possible quark rearrangements. Some detailed reviews on the $\DKM$ model can be found in Refs.~\cite{Roberts:1992js, Blundell:1996as, Li:2008mza, Li:2008we, Li:2008et}.

The transition operator $T$ of the decay  $A\rightarrow BC$ in the $\DKM$ model is given by
\begin{eqnarray}
T&=&-3\gamma\sum_m \left< 1,m;1,-m|0,0\right> \int\!\!
\ddd \bsp_3\ddd \bsp_4\delta^3(\bsp_3\!+\!\bsp_4)\nonumber\\
&&\qquad \Ycal_1^m\left(\frac{\bsp_3\!-\!\bsp_4}{2}\right
)\chi_{1\!-\!m}^{34}\phi_0^{34}\omega_0^{34}b_3^\dagger(\bsp_3)d_4^\dagger(\bsp_4),
\end{eqnarray}
where the $\gamma$ is a dimensionless parameter denoting the production strength of the quark-antiquark pair $q_3\bar{q}_4$ with quantum number $J^{PC}=0^{++}$. $\bsp_3$ and  $\bsp_4$ are the momenta of the created quark  $q_3$ and  antiquark $\bar{q}_4$, respectively. $\chi^{34}_{1,-m}$, $\phi^{34}_0$, and $\omega^{34}_0$ are the spin, flavor, and color wave functions of $q_3\bar{q}_4$, respectively. The solid harmonic polynomial  ${\Ycal}^m_1(\bsp)\equiv|\bsp|^1Y^m_1(\theta_p, \phi_p)$ reflects the momentum-space distribution of the $q_3\bar{q_4}$.

The $S$ matrix of the process $A\rightarrow BC$ is defined by
\begin{eqnarray}
\langle BC|S|A\rangle=I-2\pi i\delta(E_A-E_B-E_C)\langle BC|T|A\rangle,
\end{eqnarray}
where $|A\rangle$ ($|B\rangle$,$|C\rangle$) is the mock meson defined by ~\cite{Hayne:1981zy}
\begin{eqnarray}
&&|A(n^{2S_A+1}_AL_{A}\,\mbox{}_{J_A M_{J_A}})(\bsp_A)\rangle
\equiv \nonumber\\
&& \sqrt{2E_A}\sum_{M_{L_A},M_{S_A}}\langle L_A M_{L_A} S_A
M_{S_A}|J_A
M_{J_A}\rangle\nonumber\\
&&\times  \int d^3\bsp_A\psi_{n_AL_AM_{L_A}}(\bsp_A)\chi^{12}_{S_AM_{S_A}}
\phi^{12}_A\omega^{12}_A\nonumber\\
&&\times  \left|q_1\left({\scriptstyle
\frac{m_1}{m_1+m_2}}\bsp_A+\bsp_A\right)\bar{q}_2
\left({\scriptstyle\frac{m_2}{m_1+m_2}}\bsp_A-\bsp_A\right)\right\rangle.
\end{eqnarray}
Here, $m_1$ and $m_2$ ($\bsp_1$ and $\bsp_2$) are the masses (momenta) of the
quark $q_1$ and the antiquark $\bar{q}_2$, respectively; $\bsp_A=\bsp_1+\bsp_2$,
$\bsp_A=\frac{m_2\bsp_1-m_1\bsp_2}{m_1+m_2}$;
$\chi^{12}_{S_AM_{S_A}}$, $\phi^{12}_A$, $\omega^{12}_A$, and
$\psi_{n_AL_AM_{L_A}}(\bsp_A)$ are the spin, flavor, color, and
space wave functions of the meson $A$ composed of $q_1\bar{q}_2$ with total energy $E_A$, respectively. $n_A$ is the radial quantum number of the meson $A$. $\boldsymbol{S}_A=\boldsymbol{s}_{q_1}+\boldsymbol{s}_{\bar{q}_2}$, $\boldsymbol{J}_A=\boldsymbol{L}_A+\boldsymbol{S}_A$, $\boldsymbol{s}_{q_1}(\boldsymbol{s}_{\bar{q}_2})$ is the spin of $q_1(\bar{q}_2)$, and $\boldsymbol{L}_A$ is the relative orbital angular momentum between $q_1$ and $\bar{q}_2$.

The transition matrix element $\langle BC|T|A\rangle$ can be written as
\begin{eqnarray}
\langle BC|T|A\rangle=\delta^3(\bsp_A-\bsp_B-\bsp_C){\cal{M}}^{M_{J_A}M_{J_B}M_{J_C}}(\bsp),
\end{eqnarray}
where the helicity amplitude ${\Mcal}^{M_{J_A}M_{J_B}M_{J_C}}
(\bsp)$ is
\begin{eqnarray}
&&{\Mcal}^{M_{J_A}M_{J_B}M_{J_C}}(\bsp)=\gamma\sqrt{8E_AE_BE_C} \sum_{M_{L_A},M_{S_A}}
\nonumber\\&&\times \sum_{M_{L_B},M_{S_B}} \sum_{M_{L_C},M_{S_C}}\sum_m
\langle L_A M_{L_A} S_AM_{S_A}|J_AM_{J_A}\rangle\nonumber\\
&&\times\langle L_B M_{L_B} S_BM_{S_B}|J_BM_{J_B}\rangle\langle L_C M_{L_C} S_CM_{S_C}|J_CM_{J_C}\rangle\nonumber\\
&&\times\langle 1m1-m|00\rangle\langle \chi^{14}_{S_BM_{S_B}}\chi^{32}_{S_CM_{S_C}}|\chi^{12}_{S_AM_{S_A}}\chi^{34}_{1-m}\rangle\nonumber\\
&&\times[f_1I(\bsp,m_1,m_2,m_3)\nonumber\\
&&+(-1)^{1+S_A+S_B+S_C}f_2I(-\bsp,m_2,m_1,m_3)]
\label{helicity}
\end{eqnarray}
with $f_1=\langle \phi^{14}_B\phi^{32}_C|\phi^{12}_A\phi^{34}_0\rangle$ and $f_2=\langle \phi^{32}_B\phi^{14}_C|\phi^{12}_A\phi^{34}_0\rangle$, and
\begin{eqnarray}
I(\bsp,m_1,m_2,m_3)=&&\int d^3\bsp\psi^*_{n_BL_BM_{L_B}}\left({\scriptstyle
\frac{m_3}{m_1+m_3}}\bsp_B+\bsp\right)\nonumber\\&&\times\psi^*_{n_CL_CM_{L_C}}\left({\scriptstyle
\frac{m_3}{m_2+m_3}}\bsp_B+\bsp\right)\nonumber\\&&\times\psi_{n_AL_AM_{L_A}}\left(\bsp_B+\bsp\right){\Ycal}^m_1(\bsp),
\label{overlap space}
\end{eqnarray}
where $\bsp={\bsp}_B=-{\bsp}_C$, $\bsp=\bsp_3$, $m_3$ is the mass of the created quark $q_3$. Also, the helicity amplitude can be transform into the partial wave amplitude ${\Mcal}^{LS}(\bsp)$~\cite{Jacob:1959at},
\begin{eqnarray}
{\Mcal}^{LS}(\bsp)&=&
\sum_{\renewcommand{\arraystretch}{.5}\begin{array}[t]{l}
\scriptstyle M_{J_B},M_{J_C},\\\scriptstyle M_S,M_L
\end{array}}\renewcommand{\arraystretch}{1}\!\!
\langle LM_LSM_S|J_AM_{J_A}\rangle \nonumber\\
&&\langle
J_BM_{J_B}J_CM_{J_C}|SM_S\rangle\nonumber\\
&&\times\int
d\Omega\,\mbox{}Y^\ast_{LM_L}{\Mcal}^{M_{J_A}M_{J_B}M_{J_C}}
(\bsp). \label{pwave}
\end{eqnarray}

Various $\DKM$ models exist in literature and typically differ in the choices of the pair-production vertex, the phase space conventions, and the meson wave functions employed. In this work, we restrict to the simplest vertex as introduced originally by Micu~\cite{Micu:1968mk} which assumes a spatially constant pair creation strength $\gamma$ for the $u\bar u$ and $d \bar d$ pairs. For the $s \bar s$ pair, the creation strength is multiplied by a factor $m_u/m_s$. The wave functions can be obtained from the modified relativized quark model including the screening effects. With the relativistic phase space, the decay width
$\Gamma(A\rightarrow BC)$ can be expressed in terms of the partial wave amplitude
\begin{eqnarray}
\Gamma(A\rightarrow BC)= \frac{\pi
|\bsp|}{4M^2_A}\sum_{LS}|{\Mcal}^{LS}(\bsp)|^2, \label{width1}
\end{eqnarray}
where $|\bsp|=\sqrt{[M^2_A-(M_B+M_C)^2][M^2_A-(M_B-M_C)^2]}/{2M_A}$,
and $M_A$, $M_B$, and $M_C$ are the masses of the mesons $A$, $B$,
and $C$, respectively.

\section{CALCULATION AND RESULTS}
\label{sec:result}

\subsection{ Mass spectrum}

\begin{table}[htpb]
	\begin{center}
		\caption{ \label{tab:parameter}Parameters in the Godfrey-Isgur's relativized quark model \cite{Godfrey:1985xj}.  }
		\begin{tabular}{cccccc}
			\hline\hline
			Parameter  & value & Parameter       &value     & Parameter              &value\\
			\hline
			$m_u$~(GeV) &0.22   &$b$~(GeV$^2$)    &0.18      &$\epsilon_{\rm c}$      &-0.168 \\
			$m_d$~(GeV) &0.22   &$c$~(GeV)        &-0.253    &$\epsilon_{\rm t}$      &+0.025\\
			$m_s$~(GeV) &0.419  &$\sigma_0$~(GeV) &1.8       &$\epsilon_{\rm{so(v)}}$  &-0.035\\
			$m_c$~(GeV) &1.628  &$s$              &1.55      &$\epsilon_{\rm so(s)}$  &+0.055\\
			\hline\hline
		\end{tabular}
	\end{center}
\end{table}

\begin{table*}[htpb]
	\begin{center}
		\caption{\label{tab:masses}Comparison of the experimental data and theoretical results with different $\mu$. We take $\mu$=0.04, 0.045, 0.05 GeV to show the results with the modified relativized quark model with screened potential. We also list the $\chi^{2}$ values for different models. }
		\footnotesize
		\begin{tabular}{cccccccc}
			\hline\hline
			$$                          &$n^{2s+1}L_{J}$   &Experimental values   &GI model       &$$  &Modified GI model  &$$   &$$        \\
			$$                          &$$                   &$$                 &$$           &$\mu=0.04$  &$\mu=0.045$  &$\mu=0.05$        \\
			\hline
			$D_{s}^{\pm}$                 &$1^{1}S_{0}$   &1968.34$\pm$0.07       &1979          &1968       &1968      &1968                  \\
			$D_{s}^{\ast\pm}$             &$1^{3}S_{1}$   &2112.2$\pm$0.4         &2129          &2114       &2114      &2113                 \\
			$D_{s2}^{\ast}(2573)$         &$1^{3}P_{2}$   &2569.1$\pm$0.8         &2592          &2559       &2556      &2553                 \\
			$D_{s1}^{\ast}(2700)^{\pm}$   &$2^{3}S_{1}$   &2708$^{+4.0}_{-3.4}$             &2732          &2681       &2675      &2670                  \\
			$D_{s1}^{\ast}(2860)^{\pm}$   &$1^{3}D_{1}$   &2859$\pm$12$\pm$24     &2899          &2839       &2833      &2827                 \\
			$D_{s3}^{\ast}(2860)^{\pm}$   &$1^{3}D_{3}$   &2860.5$\pm$2.6$\pm$6.5 &2917          &2858       &2852      &2846                 \\
			$\chi^{2}$                     &$$ &$$                                &666          &55.17      &86.23     &119.39             \\
			\hline\hline
		\end{tabular}
	\end{center}
\end{table*}

\begin{figure*}[htbp]
	\begin{center}
		\includegraphics[scale=0.4]{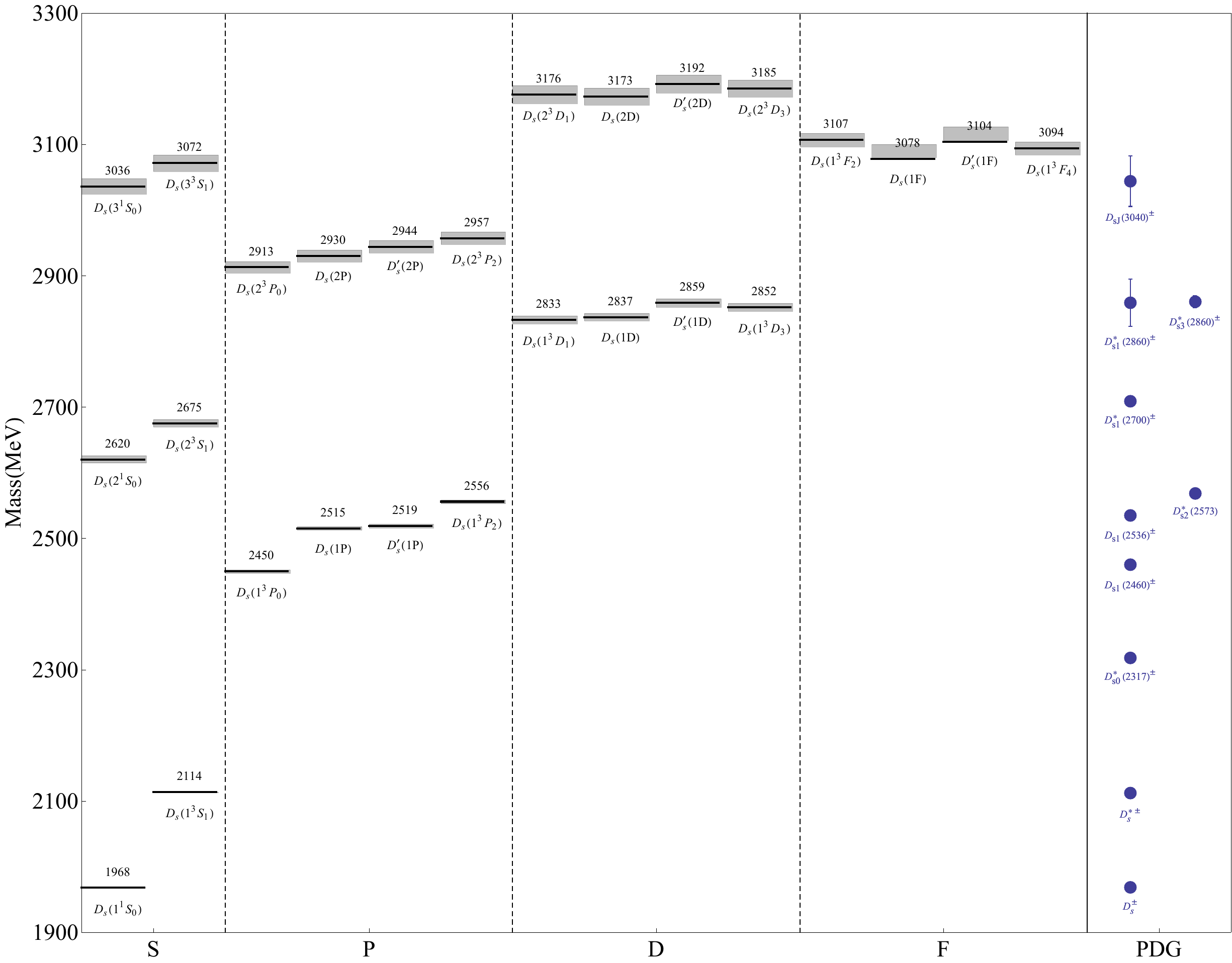}
		\vspace{0.0cm}\caption{Mass spectrum of the charmed-strange mesons in units of MeV. The black lines show the MGI model with $\mu=0.045$ GeV and the shaded regions stand for the theoretical uncertainties with $\mu=0.04-0.05$ GeV. The dark blue dot denote the experimental data~\cite{Zyla:2020zbs} and the vertical lines represent the errors.}\label{fig:spectrum}
	\end{center}
\end{figure*}

The relevant parameters used in the original relativized quark model are listed in Table~\ref{tab:parameter}~\cite{Godfrey:1985xj}. When the screening effects are included, and extra parameter $\mu$ is introduced, which reflects the strength of screening effects. In present work, we can get the parameter $\mu$ by reproducing the experimental data of low-lying states. As mentioned in the Introduction, seven states, $D_s$, $D_s^*$, $D_{s1}(2536)$, $D_{s2}^*(2573)$, $D_{s1}^*(2700)$, $D_{s1}^*(2860)$, and $D_{s3}^*(2860)$, can be reasonably classified in the conventional charmed-strange mesons. Since the $D_{s1}(2536)$ is a mixture of the $D_s(1^1P_1)$ and $D_s(1^3P_1)$ states, we do not include it when determining the parameter $\mu$ . Also, the overall constant $c$ is readjusted by fixing the mass of $D_s(1^1S_0)$ to 1968 MeV when the $\mu$ varies.

\begin{table*}[htbp]
	\begin{center}
		\caption{Our predicted masses of charmed-strange mesons compared with the experimental data and other quark model predictions. The mixing angles of $D_{s}$--$D_{s}^{\prime}$ obtained in this work are $\theta_{1P}$=$-42.7^{\circ}$, $\theta_{2P}$=$-31.4^{\circ}$, $\theta_{1D}$=$-39.4^{\circ}$, $\theta_{2D}$=$-38.4^{\circ}$, $\theta_{1F}$=$-39.9^{\circ}$. Horizontal lines indicate that the corresponding masses are not calculated in the references. The units are in MeV.}
		\label{tab:mass}
		\begin{tabular}{ c c c c c c c c c c c c }
			\hline\hline
			State &$J^{P}$ &Ours  &NLZ~\cite{Ni:2021pce}  &EFG~\cite{Ebert:2009ua}  &ZVR~\cite{Zeng:1994vj}  &GM~\cite{Godfrey:2015dva} &LNR~\cite{Lahde:1999ih}     &DE~\cite{DiPierro:2001dwf}     &LJM~\cite{Li:2010vx}     &GI~\cite{Godfrey:1985xj} &Exp~\cite{Zyla:2020zbs}  \\
			\hline	
			$D_{s}(1^{1}S_{0})$        &$0^{-}$     &1968  &1969        &1969 &1940 &1979 &1975     &1965   &1969    &1979     &1968.34$\pm$0.07            \\	
			$D_{s}(1^{3}S_{1})$        &$1^{-}$     &2114  &2112         &2111 &2130 &2129 &2180     &2113   &2107    &2129   &2112.2$\pm$0.4             \\	
			$D_{s}(2^{1}S_{0})$        &$0^{-}$     &2620  &2649         &2688 &2610 &2673 &2659     &2700   &2640    &2673  & 2591$\pm$6$\pm$7              \\	
            $D_{s}(2^{3}S_{1})$        &$1^{-}$     &2675  &2737    &2731 &2730  &2732 &2722     &2806   &2714    &2732         &2708$^{+4.0}_{-3.4}$        \\	
            $D_{s}(3^{1}S_{0})$        &$0^{-}$     &3072  &3126                      &3219 &3090 &3154 &3044     &3259   &-      &-          & -      \\	
            $D_{s}(3^{3}S_{1})$        &$1^{-}$     &3036  &3196                       &3242 &3190  &3193  &3087     &3345   &-      &-          & -      \\	
            \hline	
            $D_{s}(1^{3}P_{0})$        &$0^{+}$     &2450  &2409          &2509 &2380 &2484 &2455     &2487   &2344    &2484        &2317.8$\pm$0.5       \\	
            $D_{s}(1P)$                &$1^{+}$     &2515  &2528                       &2536 &2510 &2549 &2502     &2535   &2488    &-  & -         \\	
            $D_{s}(1P^{\prime})$       &$1^{+}$     &2519  &2545                      &2574 &2520  &2556 &2522    &2605   &2510    &-     & -        \\	
            $D_{s}(1^{3}P_{2})$        &$2^{+}$     &2556  &2575          &2571 &2580 &2592 &2586     &2581   &2559    &2592         &2569.1$\pm$0.8  \\	
            $D_{s}(2^{3}P_{0})$        &$0^{+}$     &2913  &2940                      &3054 &2900 &3005 &2901     &3067   &2830    &3005           \\	
            $D_{s}(2P)$                &$1^{+}$     &2930  &3002                     &3067 &3000  &3018 &2928     &3114   &2958    &-     & -     \\	
            $D_{s}(2P^{\prime})$       &$1^{+}$     &2944  &3026                     &3154 &3010  &3038 &2942    &3165   &2995    &-        & -   \\	
            $D_{s}(2^{3}P_{2})$        &$2^{+}$     &2957  &3053                       &3142 &3060 &3048 &2988     &3157   &3040    &3048    & -    \\	
            \hline	
            $D_{s}(1^{3}D_{1})$        &$1^{-}$     &2833  &2843       &2913 &2820  &2899 &2845     &2913   &2804    &2899    &2859$\pm$12$\pm$24      \\	
            $D_{s}(1D)$                &$2^{-}$     &2837  &2851                      &2931 &2860   &2900 &2838    &2900   &2788    &-        & -  \\	
            $D_{s}(1D^{\prime})$       &$2^{-}$     &2859  &2911                      &2961 &2880 &2926 &2856     &2953   &2849    &-     & -   \\	
            $D_{s}(1^{3}D_{3})$        &$3^{-}$     &2852  &2882  &2971 &2900 &2917 &2857     &2925   &2811    &2917       &2860.5$\pm$2.6$\pm$6.5    \\	
            $D_{s}(2^{3}D_{1})$        &$1^{-}$     &3176  &3233                       &3383 &3250 &3306 &3172     &-     &3217    &3306    & -      \\
            $D_{s}(2D)$                &$2^{-}$     &3173  &3267                      &3403 &3280 &3298 &3144     &-     &3217    &-         & -   \\	
            $D_{s}(2D^{\prime})$       &$2^{-}$     &3192  &3306                       &3456 &3290  &3323 &3167    &-     &3260    &-       & -     \\	
            $D_{s}(2^{3}D_{3})$        &$3^{-}$     &3185  &3299                      &3469 &3310 &3311 &3157     &-     &3240    &3311     & -         \\	
            \hline	
            $D_{s}(1^{3}F_{2})$        &$2^{+}$     &3107  &3176                    &3230 &3120 &3208 & -       &3224   &-      &3208      & -        \\	
            $D_{s}(1F)$                &$3^{+}$     &3078  &3123                       &3254 &3130 &3186 & -       &-     &-      &-       & -        \\	
            $D_{s}(1F^{\prime})$       &$3^{+}$     &3104  &3205                       &3266 &3150 &3218 & -       &-     &-      &-        & -       \\	
            $D_{s}(1^{3}F_{4})$        &$4^{+}$     &3094  &3134                       &3300 &3160 &3190 &-       &3220   &-      &3190    & -        \\	
			\hline\hline
		\end{tabular}
	\end{center}
\end{table*}

The mass spectrum of the charmed-strange meson with $\mu$ from 0.04 to 0.05 GeV is listed in Table~\ref{tab:masses}. For comparison, the experiment data and predictions of the original relativized quark model are also presented. It can be seen that the measured masses of the low-lying states can be well reproduced and the predicted spectrum in the screened potential is improved significantly.  Moreover, we can estimate the corresponding $\chi^{2}$ and present them in Table~\ref{tab:masses}. Here the $\chi^{2}$ can be defined as
\begin{eqnarray}
\chi^2=\sum_i\left(\frac{\mathcal{A}_{\mathrm{Th}}(i)-\mathcal{A}_{\mathrm{Exp}}(i)}{\mathrm{Error}(i)}\right)^2, \label{chi2}
\end{eqnarray}
where$\mathcal{A}_{\mathrm{Th}}(i)$, $\mathcal{A}_{\mathrm{Exp}}(i)$, and $\mathrm{Error}(i)$ are theoretical values, experimental values, and experimental errors, respectively. With the reasonable range of $\mu$, the $\chi^2$ of screened potential is significantly smaller than that of original relativized quark model.

It should be mentioned that the $\chi^{2}$ is not the only criterion of the performances for different predictions in quark models. From Eq.~(\ref{chi2}), if the experimental accuracies of several states are high enough, the model with the smallest $\chi^2$ may only reproduce these few states and fail to describe the whole mass spectrum. Phenomenologically, we also expect the absolute value $|\mathcal{A}_{\mathrm{Th}}(i)-\mathcal{A}_{\mathrm{Exp}}(i)|$ for each state is not too large, such that these states can be interpreted in the conventional $c\bar s$ picture. In the range of $\mu=0.04\sim 0.05$ GeV, the results meet the above requirements.

Hence, we prefer to choose $\mu=0.045$ GeV to calculate the mass spectrum of charmed-strange mesons, and take the masses with $\mu=0.04$ and 0.05 GeV as the theoretical uncertainties. With $\mu=0.045$ GeV, one can obtain the constant $c$ equals to $-0.243$. The theoretical predictions together with experimental data are shown in Figure~\ref{fig:spectrum}. It can be seen that the $D_{s0}(2590)^+$ can be assigned as the $D_s(2^1S_0)$ state according to its mass. Moreover, we compare the predictions of different models in Table~\ref{tab:mass}, and find that they give rather different predictions for the higher states. Also, the screening effects become increasingly important as the masses go up. The information on highly excited states is crucial to distinguish these different models and test our screened potential.

\subsection{Strong decays}

Besides the mass spectrum, the strong decay behaviors are essential to clarify the internal structure of a new resonance. In this work, the $\DKM$ model is adopted to investigate the strong decays of the $D_{s0}(2590)$. While we calculate the mass spectrum, the corresponding wave functions of mesons are also obtained. Then, only one parameter $\gamma$ in the $\DKM$ model needs to determine. We can assume that the charmed-strange mesons share the same $\gamma$, and fit this parameter from the known states. Among the seven reasonably classified states, $D_s$ only decays though weak processes, $D_s^*$ has no OZI-allowed strong decay, and the strong decays of $D_{s1}(2536)$ depend on the mixing angle sensitively. Hence, we adopt the remaining resonances, $D_{s2}^*(2573)$, $D_{s1}^*(2700)$, $D_{s1}^*(2860)$, and $D_{s3}^*(2860)$, to fit the parameter $\gamma$.

According to the fitting process, the $\gamma=9.32$ is obtained, and the strong decay behaviors of the $D_{s2}^*(2573)$, $D_{s1}^*(2700)$, $D_{s1}^*(2860)$, and $D_{s3}^*(2860)$ are listed in Table~\ref{tab:Ds1}. It can be seen that the calculated widths of $D_{s2}^\ast(2573)$ and $D_{s3}^\ast(2860)$ are consistent with the experimental data within errors, and the theoretical width of $D_{s1}^\ast(2700)$ and $D_{s1}^\ast(2860)$ seems a little bit larger. These differences may arise from the theoretical uncertainties of the $\DKM$ model or the possible complicated $S-D$ mixing mechanism for the $D_{s1}^\ast(2700)$ and $D_{s1}^\ast(2860)$ states. Hence, with the $\gamma=9.32$, the strong decay behaviors of these four states are fairly described. We employ this value to investigate the strong decays of $D_{s0}(2590)$.

\begin{table}[h]
	\begin{center}
		\caption{Decay widths of $D_{s2}^{\ast}(2573)$, $D_{s1}^{\ast}(2700)^{+}$, $D_{s1}^\ast(2860)^{+}$ and $D_{s3}^\ast(2860)^{+}$ with fitted $\gamma = 9.32$ (in MeV).}
		\label{tab:Ds1}
		\begin{tabular}{ c c c c c}
			\hline\hline
			Mode                  &$D_{s2}^{\ast}(2573)$    &$D_{s1}^{\ast}(2700)^{+}$   &$D_{s1}^\ast(2860)^{+}$   &$D_{s3}^\ast(2860)^{+}$    \\
			\hline
			$DK$                         & 12.07                     & 61.13                 &150.72             &25.65   \\
			$D^{\ast}K$                  & 1.27                      &116.05                &76.36             &17.27    \\	
			$DK^{\ast}$                  & 0.03                      &3.69                 &45.22             & 1.32        \\
			$D_{s}\eta$                  & -                         &1.61                &10.78             &0.52     \\	
			$D_{s}^{\ast}\eta$           &  -                        & -                 &3.72           & 0.16      \\
			Total width                  &  13.37                    &182.48                & 286.80           &44.92        \\
			Experiment                   & $16.9\pm0.7 $             & $122\pm10$           & $159\pm23\pm77 $  & $53\pm7\pm7$  \\
			\hline\hline
		\end{tabular}
	\end{center}
\end{table}

\begin{table}[h]
	\begin{center}
		\caption{Decay widths of $D_{s0}(2590)^{+}$ as the $D_{s}(2^{1}S_{0})$ state with fitted $\gamma =9.32 $ (in MeV).}
		\label{tab:Ds11}
		\begin{tabular}{ c c }
			\hline\hline
			Mode                  &$D_{s0}(2590)^{+}$       \\
			\hline	
			$D^{\ast+}K^{0}$        & 35.52                         \\	
			$D^{\ast0}K^{+}$        & 39.38                          \\	
			Total width             &74.90                        \\
			Experiment              & $89\pm16\pm12$                    \\
			\hline\hline
		\end{tabular}
	\end{center}
\end{table}

\begin{figure}[htbp]
	\begin{center}
		\includegraphics[scale=0.4]{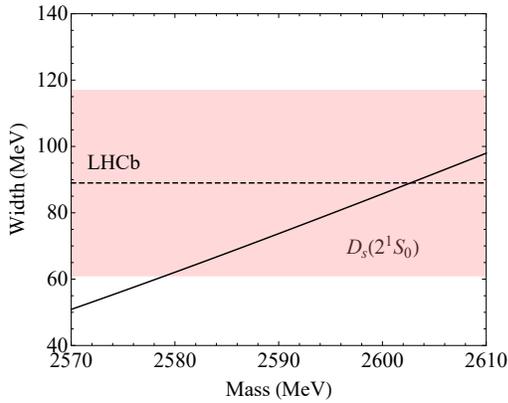}
		\caption{The dependence of the total width of $D_{s}(2^{1}S_{0})$ on the initial state mass.}\label{fig:dependent}
	\end{center}
\end{figure}

The results of $D_{s0}(2590)$ as the $D_{s}(2^{1}S_{0})$ state are listed in Table~\ref{tab:Ds11}. The calculated width is about 75 MeV, which agrees well with the experimental data $89\pm16\pm12$ MeV. Also, the dependence on the mass of initial state is shown in Fig.\ref{fig:dependent}. When the mass of initial $D_{s}(2^{1}S_{0})$ state varies from 2570 to 2610 MeV, the total width lies in the range of 51 to 98 MeV. Our results indicate that the $D_{s0}(2590)^+$ observed by LHC Collaboration can be interpreted as the conventional $D_{s}(2^{1}S_{0})$ state.

\section{SUMMARY AND CONCLUSION}

\label{sec:summary}

In this work, we investigate the mass spectrum of charmed-strange mesons with the modified relativized quark model including the screening effects. With reasonable strength of screening effects, the calculated mass spectrum can explain the $D_{s0}(2590)^{+}$ as well as other known charmed-strange mesons. The information on highly excited states is crucial to distinguish various predictions and test our results with screened potential.

Besides the mass spectrum, the strong decays of $D_{s0}(2590)^{+}$ as $D_{s}(2^{1}S_{0})$ state are also investigated in the $\DKM$ model with the obtained relativistic wave functions. The calculated width is about 75 MeV, which agrees well with the experimental data $89\pm16\pm12$ MeV. Our results indicate that the $D_{s0}(2590)^{+}$ can be interpreted as the conventional $D_{s}(2^{1}S_{0})$ state.

\section{Acknowledgements}
This work is supported by the China Postdoctoral Science Foundation Funded Project (Project No.:2021M701086), Natural Science Foundation of Henan (212300410123). Qi-Fang L\"u is supported by the National Natural Science Foundation of China under Grants No. 11705056 and No. U1832173, by the Key Project of Hunan Provincial Education Department under Grant No. 21A0039, and by the State Scholarship Fund of China Scholarship Council under Grant No. 202006725011.


\begin{thebibliography}{99}

\bibitem{BaBar:2003oey}
B.~Aubert \textit{et al.} [BaBar],
``Observation of a narrow meson decaying to $D_s^+ \pi^0$ at a mass of 2.32-GeV/c$^2$,''
Phys. Rev. Lett. \textbf{90}, 242001 (2003).

\bibitem{CLEO:2003ggt}
D.~Besson \textit{et al.} [CLEO],
``Observation of a narrow resonance of mass 2.46-GeV/c**2 decaying to D*+(s) pi0 and confirmation of the D*(sJ)(2317) state,''
Phys. Rev. D \textbf{68}, 032002 (2003).


\bibitem{Ni:2021pce}
R.~H.~Ni, Q.~Li and X.~H.~Zhong,
``Mass spectra and strong decays of charmed and charmed-strange mesons,''
[arXiv:2110.05024 [hep-ph]].



\bibitem{Ebert:2009ua}
D.~Ebert, R.~N.~Faustov and V.~O.~Galkin,
``Heavy-light meson spectroscopy and Regge trajectories in the relativistic quark model,''
Eur. Phys. J. C \textbf{66}, 197-206 (2010).



\bibitem{Zeng:1994vj}
J.~Zeng, J.~W.~Van Orden and W.~Roberts,
``Heavy mesons in a relativistic model,''
Phys. Rev. D \textbf{52}, 5229-5241 (1995).

\bibitem{Godfrey:2015dva}
S.~Godfrey and K.~Moats,
``Properties of Excited Charm and Charm-Strange Mesons,''
Phys. Rev. D \textbf{93}, 034035 (2016).

\bibitem{Lahde:1999ih}
T.~A.~Lahde, C.~J.~Nyfalt and D.~O.~Riska,
``Spectra and M1 decay widths of heavy light mesons,''
Nucl. Phys. A \textbf{674}, 141-167 (2000).



\bibitem{DiPierro:2001dwf}
M.~Di Pierro and E.~Eichten,
``Excited Heavy - Light Systems and Hadronic Transitions,''
Phys. Rev. D \textbf{64} , 114004 (2001).



\bibitem{Li:2010vx}
D.~M.~Li, P.~F.~Ji and B.~Ma,
``The newly observed open-charm states in quark model,''
Eur. Phys. J. C \textbf{71}, 1582 (2011).



\bibitem{ParticleDataGroup:2021}
Particle Data Group online (2021),
https://pdg.lbl.gov/.

\bibitem{LHCb:2020pxc}
R.~Aaij \textit{et al.} [LHCb],
``Amplitude analysis of the $B^+\to D^+D^-K^+$ decay,''
Phys. Rev. D \textbf{102}, 112003 (2020).

\bibitem{LHCb:2020bls}
R.~Aaij \textit{et al.} [LHCb],
``A model-independent study of resonant structure in $B^+\to D^+D^-K^+$ decays,''
Phys. Rev. Lett. \textbf{125}, 242001 (2020).

\bibitem{LHCb:2020gnv}
R.~Aaij \textit{et al.} [LHCb],
``Observation of a New Excited $D^+_s$ Meson in $B^0 \rightarrow D^- D^+ K^+ \pi^-$ Decays,''
Phys. Rev. Lett. \textbf{126} no.12, 122002 (2021).

\bibitem{Godfrey:1985xj}
S.~Godfrey and N.~Isgur,
``Mesons in a Relativized Quark Model with Chromodynamics,''
Phys. Rev. D \textbf{32}, 189-231 (1985).



\bibitem{Wang:2021orp}
G.~L.~Wang, W.~Li, T.~F.~Feng, Y.~L.~Wang and Y.~B.~Liu,
``The newly observed state $D_{s0}(2590)^{+}$ and width of $D^*(2007)^0$,''
[arXiv:2107.01751 [hep-ph]].

\bibitem{Xie:2021dwe}
J.~M.~Xie, M.~Z.~Liu and L.~S.~Geng,
``$D_{s0}(2590)$ as a dominant $c\bar{s}$ state with a small $D^{*}K$ component,''
Phys. Rev. D \textbf{104} no.9, 094051 (2021).

\bibitem{Ortega:2021fem}
P.~G.~Ortega, J.~Segovia, D.~R.~Entem and F.~Fernandez,
``The $D_{s0}(2590)^+$ as the dressed $c\bar s(2^1S_0)$ meson in a coupled-channels calculation,''
[arXiv:2111.00023 [hep-ph]].

\bibitem{Tan:2021bvl}
Y.~Tan and J.~Ping,
``$D^*_{s0}(2317)$ and $D_{s1}(2460)$ in an unquenched quark model,''
[arXiv:2111.04677 [hep-ph]].

\bibitem{Browder:2003fk}
T.~E.~Browder, S.~Pakvasa and A.~A.~Petrov,
``Comment on the new D(s)(*)+ pi0 resonances,''
Phys. Lett. B \textbf{578}, 365-368 (2004).

\bibitem{Hwang:2005tm}
D.~S.~Hwang and D.~W.~Kim,
``Mass shift of D/sJ(2317)* by coupled channel effect,''
J. Phys. Conf. Ser. \textbf{9}, 63-66 (2005).

\bibitem{Lu:2006ry}
J.~Lu, X.~L.~Chen, W.~Z.~Deng and S.~L.~Zhu,
``Pionic decays of D(sj)(2317), D(sj)(2460) and B(sj)(5718), B(sj)(5765),''
Phys. Rev. D \textbf{73} , 054012 (2006).

\bibitem{Bicudo:2005de}
P.~Bicudo,
``Confining quark-model suggestion against Ds*(2317) and Ds*(2460) as chiral partners of standard Ds,''
Phys. Rev. D \textbf{74}, 036008 (2006).


\bibitem{Mohler:2011ke}
D.~Mohler and R.~M.~Woloshyn,
``$D$ and $D_s$ meson spectroscopy,''
Phys. Rev. D \textbf{84} , 054505 (2011).

\bibitem{MartinezTorres:2011pr}
A.~Martinez Torres, L.~R.~Dai, C.~Koren, D.~Jido and E.~Oset,
``The $KD$, $\eta D_s$ interaction in finite volume and the nature of the $D_{s^* 0}(2317)$ resonance,''
Phys. Rev. D \textbf{85}, 014027 (2012).

\bibitem{Mohler:2013rwa}
D.~Mohler, C.~B.~Lang, L.~Leskovec, S.~Prelovsek and R.~M.~Woloshyn,
``$D_{s0}^*(2317)$ Meson and $D$-Meson-Kaon Scattering from Lattice QCD,''
Phys. Rev. Lett. \textbf{111}no.22, 222001 (2013).

\bibitem{Lang:2014yfa}
C.~B.~Lang, L.~Leskovec, D.~Mohler, S.~Prelovsek and R.~M.~Woloshyn,
``Ds mesons with DK and D*K scattering near threshold,''
Phys. Rev. D \textbf{90}  no.3, 034510 (2014).

\bibitem{Ortega:2016mms}
P.~G.~Ortega, J.~Segovia, D.~R.~Entem and F.~Fernandez,
``Molecular components in P-wave charmed-strange mesons,''
Phys. Rev. D \textbf{94}no.7, 074037 (2016).

\bibitem{MartinezTorres:2017bdo}
A.~Mart\'\i{}nez Torres, E.~Oset, S.~Prelovsek and A.~Ramos,
``An analysis of the Lattice QCD spectra for $D^*_{s0}(2317)$ and $D^*_{s1}(2460)$,''
PoS Hadron2017, 024 (2018).


\bibitem{Bali:2017pdv}
G.~S.~Bali, S.~Collins, A.~Cox and A.~Sch\"afer,
``Masses and decay constants of the $D_{s0}^*(2317)$ and $D_{s1}(2460)$ from $N_f=2$ lattice QCD close to the physical point,''
Phys. Rev. D \textbf{96}no.7, 074501 (2017).

\bibitem{Albaladejo:2018mhb}
M.~Albaladejo, P.~Fernandez-Soler, J.~Nieves and P.~G.~Ortega,
``Contribution of constituent quark model $c\bar{s}$ states to the dynamics of the $D_{s0}^*(2317)$ and $D_{s1}(2460)$ resonances,''
Eur. Phys. J. C \textbf{78}  no.9, 722 (2018).

\bibitem{LHCb:2019juy}
R.~Aaij \textit{et al.} [LHCb],
``Determination of quantum numbers for several excited charmed mesons observed in $B^- \to D^{*+} \pi^- \pi^-$ decays,''
Phys. Rev. D \textbf{101}, no.3, 032005 (2020).

\bibitem{Godfrey:2014fga}
S.~Godfrey and K.~Moats,
``The $D_{sJ}^*(2860)$ Mesons as Excited D-wave $c\bar{s}$ States,''
Phys. Rev. D \textbf{90}no.11, 117501 (2014).


\bibitem{Li:2021qod}
Y.~S.~Li, X.~Liu and F.~S.~Yu,
``Revisiting semileptonic decays of \ensuremath{\Lambda}b(c) supported by baryon spectroscopy,''
Phys. Rev. D \textbf{104}no.1, 013005 (2021).

\bibitem{Godfrey:2004ya}
S.~Godfrey,
``Spectroscopy of $B_c$ mesons in the relativized quark model,''
Phys. Rev. D \textbf{70}, 054017 (2004).

\bibitem{Capstick:1985xss}
S.~Capstick and N.~Isgur,
``Baryons in a Relativized Quark Model with Chromodynamics,''
Phys. Rev. D \textbf{34}, no.9, 2809-2835 (1986)


\bibitem{Sun:2014wea}
Y.~Sun, Q.~T.~Song, D.~Y.~Chen, X.~Liu and S.~L.~Zhu,
``Higher bottom and bottom-strange mesons,''
Phys. Rev. D \textbf{89}, no.5, 054026 (2014).

\bibitem{Godfrey:2016nwn}
S.~Godfrey, K.~Moats and E.~S.~Swanson,
``$B$ and $B_s$ Meson Spectroscopy,''
Phys. Rev. D \textbf{94}, no.5, 054025 (2016).

\bibitem{Barnes:2005pb}
T.~Barnes, S.~Godfrey, and E.~S.~Swanson,
Higher charmonia,
Phys.\ Rev.\ D {\bf 72}, 054026 (2005).


\bibitem{Lu:2021kut}
Q.~F.~L\"u, D.~Y.~Chen, Y.~B.~Dong and E.~Santopinto,
``Triply-heavy tetraquarks in an extended relativized quark model,''
Phys. Rev. D \textbf{104} no.5, 054026 (2021).

\bibitem{Lu:2020cns}
Q.~F.~L\"u, D.~Y.~Chen and Y.~B.~Dong,
``Masses of fully heavy tetraquarks $QQ {\bar{Q}} {\bar{Q}}$ in an extended relativized quark model,''
Eur. Phys. J. C \textbf{80} no.9, 871  (2020).

\bibitem{Lu:2020rog}
Q.~F.~L\"u, D.~Y.~Chen and Y.~B.~Dong,
``Masses of doubly heavy tetraquarks $T_{QQ^\prime}$ in a relativized quark model,''
Phys. Rev. D \textbf{102}, no.3, 034012 (2020).


\bibitem{Lu:2020qmp}
Q.~F.~L\"u, D.~Y.~Chen and Y.~B.~Dong,
``Open charm and bottom tetraquarks in an extended relativized quark model,''
Phys. Rev. D \textbf{102} no.7, 074021 (2020).


\bibitem{Lu:2016cwr}
Q.~F.~L\"u and Y.~B.~Dong,
``X(4140) , X(4274) , X(4500) , and X(4700) in the relativized quark model,''
Phys. Rev. D \textbf{94}  no.7, 074007 (2016).

\bibitem{Lu:2016zhe}
Q.~F.~L\"u and Y.~B.~Dong,
``Masses of open charm and bottom tetraquark states in a relativized quark model,''
Phys. Rev. D \textbf{94} no.9, 094041 (2016) .

\bibitem{Lu:2019ira}
Q.~F.~L\"u, K.~L.~Wang and Y.~B.~Dong,
``The $ss \bar s \bar s$ tetraquark states and the newly observed structure $X(2239)$ by BESIII Collaboration,''
Chin. Phys. C \textbf{44}  no.2, 024101 (2020).

\bibitem{Anwar:2017toa}
M.~N.~Anwar, J.~Ferretti, F.~K.~Guo, E.~Santopinto and B.~S.~Zou,
``Spectroscopy and decays of the fully-heavy tetraquarks,''
Eur. Phys. J. C \textbf{78}  no.8, 647 (2018).

\bibitem{Anwar:2018sol}
M.~N.~Anwar, J.~Ferretti and E.~Santopinto,
``Spectroscopy of the hidden-charm $[qc][\bar q \bar c]$ and $[sc][\bar s \bar c]$ tetraquarks in the relativized diquark model,''
Phys. Rev. D \textbf{98} no.9, 094015 (2018).

\bibitem{Bedolla:2019zwg}
M.~A.~Bedolla, J.~Ferretti, C.~D.~Roberts and E.~Santopinto,
``Spectrum of fully-heavy tetraquarks from a diquark+antidiquark perspective,''
Eur. Phys. J. C \textbf{80}no.11, 1004 (2020).

\bibitem{Ferretti:2020ewe}
J.~Ferretti and E.~Santopinto,
``Hidden-charm and bottom tetra- and pentaquarks with strangeness in the hadro-quarkonium and compact tetraquark models,''
JHEP \textbf{04}, 119 (2020).


\bibitem{Li:2009zu}
B.~Q.~Li and K.~T.~Chao,
``Higher Charmonia and X,Y,Z states with Screened Potential,''
Phys. Rev. D \textbf{79} , 094004 (2009).

\bibitem{Chao:1992et}
K.~T.~Chao, Y.~B.~Ding and D.~H.~Qin,
``Possible phenomenological indication for the string Coulomb term and the color screening effects in the quark - anti-quark potential,''
Commun. Theor. Phys. \textbf{18}, 321-326 (1992).


\bibitem{Ding:1993uy}
Y.~B.~Ding, K.~T.~Chao and D.~H.~Qin,
``Screened Q anti-Q potential and spectrum of heavy quarkonium,''
Chin. Phys. Lett. \textbf{10} , 460-463 (1993).


\bibitem{Song:2015fha}
Q.~T.~Song, D.~Y.~Chen, X.~Liu and T.~Matsuki,
``Higher radial and orbital excitations in the charmed meson family,''
Phys. Rev. D \textbf{92} no.7, 074011 (2015).

\bibitem{Wang:2018rjg}
J.~Z.~Wang, Z.~F.~Sun, X.~Liu and T.~Matsuki,
``Higher bottomonium zoo,''
Eur. Phys. J. C \textbf{78} no.11, 915 (2018).


\bibitem{Pang:2017dlw}
C.~Q.~Pang, J.~Z.~Wang, X.~Liu and T.~Matsuki,
``A systematic study of mass spectra and strong decay of strange mesons,''
Eur. Phys. J. C \textbf{77} no.12, 861 (2017).


\bibitem{Pang:2019ttv}
C.~Q.~Pang,
``Excited states of $\phi$ meson,''
Phys. Rev. D \textbf{99}  no.7, 074015 (2019).


\bibitem{Pang:2018gcn}
C.~Q.~Pang, Y.~R.~Wang and C.~H.~Wang,
``Prediction for $5^{++}$ mesons,''
Phys. Rev. D \textbf{99}  no.1, 014022 (2019).


\bibitem{Hao:2019fjg}
W.~Hao, G.~Y.~Wang, E.~Wang, G.~N.~Li and D.~M.~Li,
``Canonical interpretation of the $X(4140)$ state within the $^3P_0$ model,''
Eur. Phys. J. C \textbf{80}  no.7, 626 (2020).

\bibitem{Song:2015nia}
Q.~T.~Song, D.~Y.~Chen, X.~Liu and T.~Matsuki,
``Charmed-strange mesons revisited: mass spectra and strong decays,''
Phys. Rev. D \textbf{91} , 054031 (2015).

\bibitem{Barnes:2002mu}
T.~Barnes, N.~Black, and P.~R.~Page,
Strong decays of strange quarkonia,
Phys.\ Rev.\ D {\bf 68}, 054014 (2003).
				
\bibitem{Wang:2017pxm}
G.~Y.~Wang, S.~C.~Xue, G.~N.~Li, E.~Wang and D.~M.~Li,
``Strong decays of the higher isovector scalar mesons,''
Phys. Rev. D \textbf{97} no.3, 034030 (2018).
		
\bibitem{Roberts:1992js}
W.~Roberts and B.~Silvestre-Brac,
General method of calculation of any hadronic decay in the $^3P_0$ triplet model,
Few-Body Syst. 11, 171 (1992).
				
\bibitem{Blundell:1996as}
H.~G.~Blundell,
Meson properties in the quark model: A look at some outstanding problems,
hep-ph/9608473 (1996).
		
\bibitem{Barnes:1996ff}
T.~Barnes, F.~E.~Close, P.~R.~Page, and E.~S.~Swanson,
Higher quarkonia,
Phys.\ Rev.\ D {\bf 55}, 4157 (1997).
				
\bibitem{Close:2005se}
F.~E.~Close and E.~S.~Swanson,
Dynamics and decay of heavy-light hadrons,
Phys.\ Rev.\ D {\bf 72}, 094004 (2005).
					
\bibitem{Zhang:2006yj}
B.~Zhang, X.~Liu, W.~Z.~Deng, and S.~L.~Zhu,
$D_{sJ}(2860)$ and $D_{sJ}(2715)$,
Eur.\ Phys.\ J.\ C {\bf 50}, 617 (2007).
				
\bibitem{Ding:2007pc}
G.~J.~Ding and M.~L.~Yan,
$Y(2175)$: Distinguish Hybrid State from Higher Quarkonium,
Phys.\ Lett.\ B {\bf 657}, 49 (2007).
		
\bibitem{Li:2008mza}
D.~M.~Li and B.~Ma,
$X(1835)$ and $\eta(1760)$ observed by the BES Collaboration,
Phys.\ Rev.\ D {\bf 77}, 074004 (2008).
		
		
\bibitem{Li:2008we}
D.~M.~Li and B.~Ma,
The $\eta(2225)$ observed by the BES Collaboration,
Phys.\ Rev.\ D {\bf 77}, 094021 (2008).
		
\bibitem{Li:2008et}
D.~M.~Li and S.~Zhou,
Towards the assignment for the $4^1S_0$ meson nonet,
Phys.\ Rev.\ D {\bf 78}, 054013 (2008).
		
			
\bibitem{Li:2008xy}
D.~M.~Li and S.~Zhou,
Nature of the $\pi_2(1880)$
Phys.\ Rev.\ D {\bf 79}, 014014 (2009).
		
\bibitem{Li:2009rka}
D.~M.~Li and E.~Wang,
Canonical interpretation of the $\eta_2(1870)$,
Eur.\ Phys.\ J.\ C {\bf 63},297 (2009).
						
\bibitem{Li:2009qu}
D.~M.~Li and B.~Ma,
Implication of BaBar's new data on the $D_{s1}(2710)$ and $D_{sJ}(2860)$,
Phys.\ Rev.\ D {\bf 81}, 014021 (2010).
				
\bibitem{Lu:2014zua}
Q.~F.~L\"{u} and D.~M.~Li,
Understanding the charmed states recently observed by the LHCb and BaBar Collaborations in the quark model,
Phys.\ Rev.\ D {\bf 90}, 054024 (2014).
				
\bibitem{Pan:2016bac}
T.~T.~Pan, Q.~F.~L\"{u} , E.~Wang and D.~M.~Li,
Strong decays of the $X(2500)$ newly observed by the BESIII Collaboration,
Phys.\ Rev.\ D {\bf 94}, no. 5, 054030 (2016).
		
\bibitem{Lu:2016bbk}
Q.~F.~L\"{u}, T.~T.~Pan, Y.~Y.~Wang, E.~Wang, and D.~M.~Li,
Excited bottom and bottom-strange mesons in the quark model,
Phys.\ Rev.\ D {\bf 94},074012 (2016).
				

\bibitem{Hayne:1981zy}
C.~Hayne and N.~Isgur,
Phys. Rev. D \textbf{25} , 1944 (1982).


\bibitem{Jacob:1959at}
M.~Jacob and G.~C.~Wick,
On the general theory of collisions for particles with spin,
Annals Phys.\ {\bf 7}, 404 (1959).		
		
\bibitem{Micu:1968mk}
L.~Micu,
Decay rates of meson resonances in a quark model,
Nucl.\ Phys.\ B {\bf 10}, 521 (1969).
	
\bibitem{Zyla:2020zbs}
P.~A.~Zyla \textit{et al.} [Particle Data Group],
``Review of Particle Physics,''
PTEP \textbf{2020}no.8, 083C01 (2020).


\end{thebibliography}
\end{document}